# Search for the Reasons of Josephson – Like Behavior of Thin Granular Carbon Films

Sergey G.Lebedev

*Institute for Nuclear Research of Russian Academy of Sciences*
*60$^{th}$ October Anniversary Prospect 7a, 117312, Moscow, Russia*

**Abstract**

This review is described some new results in the studies of anomalous electromagnetic of thin carbon films. Starting with historical sketch of studies of superconductivity and electromagnetic in carbonaceous materials the efforts of such persons as the Nobel Prize winners Vitaly L.Ginzburg and W.A.Little, the Polish outstanding scientist K.Antonowicz, the Americans J.T.Chen, Liren Zhao and Bor Z. Jang, the German researcher P.Esquinazy and his group, the Russian scientists T.L.Makarova, V.I.Tsebro, O.E.Omel'yanovskii and A.P. Moravskii, and the Chinese scientists G.M.Zhao and Y.S.Wang has been mentioned.

The work presents the study of some new anomalous electromagnetic effects in graphite-like thin carbon films. These are:
- The fast switching ($10^{-9}$ *sec*) of electrical conductivity
- The detection of microwave radiation and its temperature dependence
- The oscillations of film stack magnetization in a magnetic field of *1-5 T*
- The generation of optical radiation during the spasmodic switching of conductivity

These effects are explained to be the consequence of carbon films granular structure, which can be considered as the Josephson junction array (*JJA*).

The description of theory of the reversed *ac* Josephson effect (*RJE*) and its application to the search of high-temperature superconducting (*SC*) phase in granular materials was done. As for magnetic properties the description of the technique of P.Barbara and co-workers, A.P.Nielsen e.a., C.Auletta et al., Mahesh Chandran and other scientists on studies of two-dimensional *JJA* has been added. The both mentioned techniques have been applied to analyze the data on electromagnetic measurements in the thin granular carbon films.

Results of atomic force microscopy (*AFM*), magnetic force microscopy (*MFM*), *dc SQUID* magnetization, *RJE*, and resistance measurements in thin carbon arc (*CA*) films are presented. The observation of a *RJE* induced voltage as well as its *RF* frequency, input amplitude, and temperature dependence reveals the existence of the Josephson-like Junction arrays. Oscillating behavior of the *dc SQUID* magnetization reminiscent of the Fraunhofer-like behavior of *SC* critical current in the range of *1-5 T* has been observed. The *dc SQUID* magnetization measurement indicates a possible elementary *102 nm SC* loop; this is compared with the *MFM* direct observations of magnetic clusters with the median size of *165 nm*. The results obtained provides a basis for non-cryogenic electronics utilizing the Josephson's effect such as contact less field-effect switcher (*FES*), the detector of $\gamma$ – radiation for registration of neutrino and dark matter, the generators and detectors of microwave radiation, the magnetic protection, the new type of micro lasers based on the magnetic vortices movement, and the low-dissipative coatings.

PACS numbers: 74.25.H; 74.76; 85.25.C. Key words: thin films; superconductivity; Josephson devices.

## I. INTRODUCTION

Physical and elecrtomagnetic properties of the carbon and its derivatives are the subjects of constant interest and extensive studies for many eyars. The discovery of fullerenes [1] has promted an upsurge of experimental and theoretical studies resulting in finding that $C_{60}$ becomes a superconductor [2] upon doping with alkali-metal atoms. In the past few years much attention has been payed to superconducting signs in the carbon single-walled nanotubes and the multiwalled nanotubes [3-4]. Recently the superconducting-like behavior at room temperature was found in highly oriented pyrolitic graphite [5-6].

The possibility of existence of superconductivity above room temperature have substantiated by the Nobel Prize winners V.L.Ginzburg [7] and W.A.Little [8]. After that many researchers with enthusiasm were started to searches of the high temperature superconductors (*HTSC*). According to the modern representations, the superconductivity is caused by coupling of separate electrons with each other's in Cooper's pairs through the chain of atoms of the crystalline lattice. Electrons as though constantly pull a chain, coordinating their movement with the partners. Thus the pair of electrons moves in a crystalline lattice as a unity and does not dissipate its energy. The greater the exchange frequency of "jerks", the more strongly the electrons are coupled in pairs and the higher the temperature of breaking-down of a superconducting state. It is noticed, that the «jerks frequency» is higher in the materials with the high melting point, such as carbon with its greater variety of chemical and structural forms. Therefore the carbon and its compounds was one of the first, which «have got under suspicion». The known Polish scientist K.Antonowicz (*1914-*

*2002*) more than *30* years ago investigated the conductive properties of the glassy carbon [9] and it's evaporated deposits [10] and has found out the effect of jump of conductivity up to three orders of magnitude. The change of conductivity was reversible, and the relaxation time made some days. Later on Antonowicz has revealed the change of samples current-voltage characteristics at the irradiation of an *Al-C-Al*-sandwich with the microwave radiation [11]. However, the change of current occurred with the time delay of about *100 minutes*, after a microwave irradiation. At the first sight such «time delay» seems to be rather strange from the point of view of electronic mechanisms. Nevertheless, Antonowicz has explained the effect observed by the superconductivity at room temperature [12].

B.Z.Jang and L.Zhao also studied the switching behaviour of carbonised materials [13]. They study the partially carbonized polyacrylonitrile fibers which were observed to undergo a resistivity change of 2 to 4 orders of magnitude at a transition temperature typically in the range of 98 °C to 200 °C. The current-voltage curves exhibited an initial supercurrent-like increase, followed by a rapid drop to a high resistance state, and then a rise in current again at a later stage. These phenomena cannot be interpreted by existing theories on switching in inorganic amorphous semiconductors. They are explainable if the microstructure of the pyrolyzed fiber is viewed as comprising nanometer-scale superconducting phases interspersed with semiconducting phases, much like a large number of Josephson junctions connected in series.

H.A.Goldberg, I.L.Kalnin, C.C.Williams, I.L.Spain have obtained the *US* patent on the carbon switching device [14]. This device was made by partially pyrolyzing polymer material by heating the material to between 500.degree. C. and 800.degree. C. Electrodes are applied to the material at two different locations to define an electrically active element therebetween. Devices made according to the teachings of the disclosure exhibit negative resistance in a portion of their voltage current domain and function as bi-directional electrical switches.

Other scientists have observed the anomalous behaviour of carbon also. G.M.Zhao and Y.S.Wang found the traces of *HTSC* with the critical temperature about *650°K* in the carbon nanotubes [4], the Russian scientists V.I.Tsebro, O.E. Omel'yanovskii and A.P.Moravskii observed the weak decay of *SC* currents in the composite made of carbon nanotubes at room temperature [3]. Paramagnetic behavior of nanostructured forms of molecular carbon $C_{60}$ has been observed by T.L.Makarova et al. [15]. The fresh arguments in support of the idea of room temperature *SC* may be the results of N.Breda et al. about the possibility the *SC* with the $T_c$~*320°K* in the samples composed of fullerene clusters $C_{28}$ [16] and also the work of K.K.Gomes et al. evidenced about the existence of electron pairing at temperatures well above of $T_c$ [17].

The author of this paper spent more than twenty years for study of the thin carbon films first as the strippers for charge particle beams. The efforts have been stressed on the radiation stability of thin carbon stripper foils, its lifetime and some structure rearrangement under ion bombarding [18-22].

Later on the attention has been payed to the electromagnetic properties of thin carbon films. Some anomalies in the electromagnetics of films formed by sputtering of spectroscopically pure graphite in electrical arc discharge (*CA* films) have been found [23-26]. This was the temperature dependent electrical resistivity jump at some critical current, the *rf*-to-*dc* conversion associated with the *ac* Josephson effect at room and some higher temperatures, the oscillations of film stack magnetization in the magnetic field of *1-5 T*, and the optical radiation emitted during the spasmodic switching of conductivity.

The carbon films studied are a composite of small (about *10-20 A*) graphite-like $sp^2$-bonding granules embedded in the matrix of amorphous carbon [27]. The conglomerate of granules connecting each other by the Josephson junctions is coupled in the many finite size loops. So this is the spin-glass like system similar to that numerically studied by C.Ebner and D. Stroud [28]. They have demonstrated the *dc*-susceptibility oscillations in Figs.1 and 2 of their paper which is very similar to that observed in the carbon films (please see Sect. III C. of this review).

Studying the conducting properties of carbon films, obtained by sputtering of the graphite in an electric arc, we have found out the jumps of electrical resistance on four-five orders of magnitude at some critical current [12]. Then, after the time delay, there came the relaxation of conductivity. This is in agreement with Antonowicz's experiments, but the results of his studies became known to us only many years later. At similarity of the investigated effects and similarity of the samples structure the direction of our researches; their results and conclusions appreciably differ from that of Antonowicz's. More detail consideration of this problem is presented in Sect. II A.

At the room temperature the critical current in our samples varies within the range of *5-500 mA* depending on type of a condensate and sample annealing conditions. With the lowering of temperature the value of a critical current increases. After a relaxation time the low resistance state was completely restored, so the samples can be used for switching repeatedly. Time of switching of such contactless switcher measured by us is about *1* nanosecond that excludes the thermal mechanism of switching.

It is possible to explain a combination of such fast switching and long relaxation time by the presence of Josephson's vortices discovered in due time by the Nobel prize winner A.A.Abrikosov working now in the Argonne National Laboratory, *USA*. These vortices represent the cylindrical objects formed by the circles of the superconducting currents inside of which there is a kernel of a normal phase with the destroyed superconductivity. Each vortice bears in itself the one quantum of the magnetic flux. The vortices get into a film through the boundary from the outside and can migrate under applied electric and magnetic fields, and also "to be hooked" by every possible defects and heterogeneity which always exist inside the film. Conditions of vortice penetration depend on the value of magnetic and the electric fields. The greater the value of a magnetic field, the less the size of formed vortices and the easier they get and move in a film. The enclosed electric field "pushes out" vortices from the film. Therefore long relaxation time of conductivity after switching can be connected with slow penetration of vortices in the film. At the same time biasing the sample with the high enough electric field will neutralize the influence of fixing barriers and will force the vortices to leave a film quickly. Actually the movement of vortices under enclosed voltage defines the high conductivity of a carbon film.

Other interesting feature of carbon films is the occurrence of a constant voltage on the contacts at the microwave irradiation, i.e. the detection of the microwave radiation. If the current is below the critical value then the Cooper's pairs can tunnel from the one superconductor to another, practically without breaking, and the Josephson's contact (*JC*) behaves as a superconductor. In another words, if a current is below the critical value the voltage on the contacts is absent. But when the current reaches the critical value, the Cooper's pair's breaks in a layer between the two superconductors. Desintegration of each pair is connected with the emission of photon which the frequency $v$ depending on the electron coupling energy according to the relation $E_b = \hbar v$, where $\hbar$ – is the Planck's constant. Such a process refers to as the non-stationary Josephson's effect and it explains the emission of the optical radiation from a *JC*. It is known also that the reversed *ac* Josephson's effect is the inducing of constant voltage on a *JC* under the microwave irradiation. The *RJE* is actively used at the research of both single *JC*, and their associations - the Josephson's media (*JM*). *30* years ago in the Antonowicz's time the representations about *RJE* and *JM* have not been developed yet. The simple description of the RJE application for the verification of HTSC is presented in Sect. II A.

The detecting of microwave in the carbon film samples – is only one illustration of *JM* reality. The other evidences of Josephson's behaviour are shown in their magnetic properties. The measurement of magnetization of the samples with the small fraction of the superconducting phase is the very labor-consuming problem, which can be solved only with the help of such high-sensitivity devices as *SQUID* (Superconducting Quantum Interference Device)-magnetometer. The design of this device is based on an interference of the weak magnetic fluxes in the sample with known magnetic flux in the *SQUID* superconducting ring. In fact the Josephson's interference allows to measure the values of the magnetic fluxes, comparable with the magnetic flux quantum $\Phi_0 = 2\cdot10^{-7}$ *Hauss·cm²*. This value has the dimension of the magnetic field by the area. If the area of a vortice makes *1 cm²* it bears the magnetic field of $2\cdot10^{-7}$ *Hauss*.

By means of *SQUID*-magnetometr we have found out the magnetization oscillations of the sample in the magnetic fields of $10^4$-$5\cdot10^4$ *Hauss* [26]. The value of the magnetic field corresponding to jumps of magnetization, and also their amplitude depend on the temperature. Each oscillation is connected with the increase of a magnetic flux by one magnetic flux quantum $\Phi_0$ in the cluster. Using the data of measurements, we have defined the average size of magnetic clusters being about *0.1 microns*. In the *MFM* we have seen the magnetic clusters and we have defined their average size of about *0.16 microns*. This is good enough agreement with the cluster sizes found out in the *SQUID*-measurements. Having compared the *MFM* and the *AFM* topological pictures, we have noticed, that, at least, some magnetic clusters coincide with the topological ones.

This review presents the results of *dc* magnetization, atomic and magnetic force microscope, low and high temperature electrical resistance measurements, *RJE* voltage, and generation of *IR* radiation under switching of conductivity in the carbon films. Each of these results cannot be explained unambiguously, but considered together they suggest the existence of *SC* phase or fluctuation in the granular carbon films. On the basis of results obtained the granular carbon films can be used in the applications as non-cryogenic Josephson devices.

## II.    *JJA* VERIFICATION CONSEPTS

## A. REVERSED AC JOSEPHSON EFFECT

A superconductor consisting of grains may be considered to be composed of coupled Josephson junctions. In *dc* measurements, the zero-resistance state can be observed if the bias current is less than the smallest critical *dc* Josephson current of all the junctions in the sample.

Consequently, a different technique is required to verify the superconducting transition when it does not result in a zero-resistance state. If some of the junctions are in the finite-voltage state, the *ac* Josephson effect may arise as *ac* current is applied to these junctions. J.T.Chen et al. had used this technique to verify the superconductivity in *Y-Ba-Cu-O* compound at the critical temperature of *240 $^oK$* [29]. This technique is discussed in detail below.

The reversed *ac* Josephson effect represents the inducing of the *dc* voltage under microwave irradiation. The behavior of the induced constant voltage associated with the *ac* Josephson effect is distinctly different from a nonvanishing, time-averaged voltage due to rectification effects associated with the asymmetrical current-voltage characteristics. Thus, it is possible to distinguish these two effects by careful and thorough examination of the induced voltage as a function of temperature, *RF* frequency, *RF* amplitude and time. The induced *dc* voltage varies in a random, oscillatory manner as a function of temperature since it is a result of a series combination of a large number of individual junctions whose individual induced *dc* voltages also vary in some random oscillatory manner as a function of frequency and amplitude.

The *RJE* has been observed in single junctions [30-32] and in arrays [33] under microwave radiation generating induced *dc* voltages as large as 0.5 *mV* in single junction [31] and on the order of volts in multiple junctions [33]. The *RF* induced *dc* voltage in these measurements ranged from a few microvolts to millivolts with its polarity changing as a function of temperature, *RF* amplitude, and *RF* frequency in a random fashion. We have observed the similar behavior in granular carbon films at room and some higher temperatures (please see the Sect. III B.).

In the pressure of the *RF* current of frequency *f*, the individual Josephson junctions in granular materials can produce the quantized *dc* voltages, $V_j=nhf/2e$, on the order of nanovolts, even in the unbiased sample. Due to the thermal smearing, the individual quantum voltages are not observable. However the *dc* voltages of the order of millivolts can be observed resulting from the summation of thousands of Josephson junctions with values of *n* as large as *100*.

In the experimental measurements of *RJE* care should be taken to allow the sample to reach the thermal equilibrium at all temperatures since the thermal emf of the order of *10 µV* or some more can easily develop in these materials. Typically, one trace of temperature dependence was taken over a period of *3 houres* or some longer. In addition, small *RF* amplitude is preferred to avoid other rectification effects. Examination of the voltage versus time on an osciloscope is especially helpful in order to separate the constant voltage associated with the reversed *ac* Josephson effect from the time-averaged voltages arising from rectification. The principal test that can distinguish the *RJE* from the more familar rectification effect is the change of the palarity of $V_{dc}$ versus from both the frequency and the amplitude of *ac* signal. It will be impossilbe to observe such kind of behavior in the frame of rectification scenario.

## B. REENTRANT AC MAGNETIC SUCCEPTIBILITY

The Meissner effect of the granular *HTSC* is known to be incomplete even for magnetic field $H<H_c$. This incompleteness is explained by flux pinning and/or to the weak-link structure of the granular samples. In order to study this behavior a *SQUID* – magnetometer is used [34]. The most striking result of this research is the observation of a paramagnetic susceptibility below $T_c$: in certain granular samples the field-cooled susceptibility not only decreases with decreasing field, but even becomes paramagnetic below $T_c$ in very low fields [34,35]. Also this effect was observed in strong (*3-7 T*) magnetic field [36]. Such kind of behavior is refered to as the "paramagnetic Meissner effect" (*PME*), although the use of the expression "Meissner effect", which is commonly associated with flux expulsion, may probably be somewhat confusing in this context.

In some cases the *PME* has been attributed to the presence of so-called "π junctions" between the grains [34]. In these junctions, the Cooper pair acquires a phase shift π across the junction, giving rise to Josephson currents, which are negative relative to conventional junctions [37]. Such a phase shift might result from magnetic impurities between the grains or non-*s*-wave pairing symmetry [38-39].

P.Barbara and co-workers, A.P.Nielsen et al, Mahesh Chandran, C. Auletta et al. [35, 40-42] have studied the elecrtomagnetics of two-dimensional *JJA* of $Nb/Al/AlO_2/Nb$ by two ways: measuring of the complex magnetic susceptibility vs temperature and magnetic field from two-dimensional array of conventional josephson junctions and calculating the magnetic susceptibility of single loop based on simple model containing of four junctions of size of *46 µm*. In the *ac* susceptibility measurements up to the magnetic field $h_{ac}$ of about *50 mOe (0.4*

*amp/m*), which corresponds to *5Φ₀* per elementary loop, the diamagnetic signal has been observed. Upon increasing $h_{ac}$ above *50 mOe (0.4 amp/m)* the paramagnetic signal reappears. As it has been shown this non-monotonic behavior of magnetization is the consequence of a magnetic field dependence of critical current in the elementary loop (the so-called Fraunhofer pattern). So the result obtained is the observation of magnetization oscillations around zero axes. At changing of magnetic field the paramagnetic responce continuously decreases passing through zero and than tourn to be diamagnetic. This behavior is repeated few times with further field increasing. Each zero axis crossover corresponds to magnetic flux entering the loop. So this is the oscillating reentrant behavior of magnetization like the results of C.Ebner and D.Stroud [28] and the behavior of carbon films. The difference is that the oscillation picture in our experiments and also of P.Barbara and co-workers, A.P.Nielsen e.a, Mahesh Chandran, C. Auletta et al. [35, 40-42] was strongly dependent on the temperature.

### III. STRANGE ELECTROMAGNETICS IN CARBON FILMS

#### A. SWITCHING OF CONDUCTIVITY

A typical experimental sample is a carbon film with the thickness of *1000-2000* Angstrom, with the sizes in plane of *1? 0.5 cm²* though these sizes can be considerably reduced without appreciable influence on the results of the experiment [23-24]. The *CA* films for our measurements were prepared by arc evaporation of *99.999%* purity carbon onto quartz substrates at room temperature [23]. Potassium-oleate ($C_{18}H_{33}O_2K$) was used as the release agent. Self-supporting films were floated from the substrate using the distilled water. The *CA* films were annealed in vacuum furnace at *1000 C* for *10 hours* [23]. Later the carbon films produced by the chemical vapor deposition (*CVD*) method were also used. The thickness of the annealed films was about *960 nm*. 2 *MeV* $H^+$-*PIXE* analysis indicated that the *Fe* concentration in the annealed films was *185 ± 38 ppm*, and the film density was about *2.25 g/cm²*. Typical film sapmple was *1 cm* long and *0.5 cm* wide.

To provide the four-probe electrical resistance measurements, thin gold wire electrical contacts were attached to the film samples with the help of silver paste. Usually contacts had dimensions of *3 mm* by *1 mm*. Also the mechanical contacts were used.

At the study of conductive properties of some carbon condensates the phenomenon of spasmodic increase in the electrical resistance on *~4-5* orders of magnitude under electric current increase up to some critical value is revealed. Some typical example of the observed behavior is shown in Fig.1. Such kind of behavior reminds the so-called field effect swithers (*FES*). The critical current decreases with temperature and at a room temperature has the values of *5 - 500 mA* (depending on deposition condition of a film) at the applied *dc* voltage in the range of *5-50 V*.

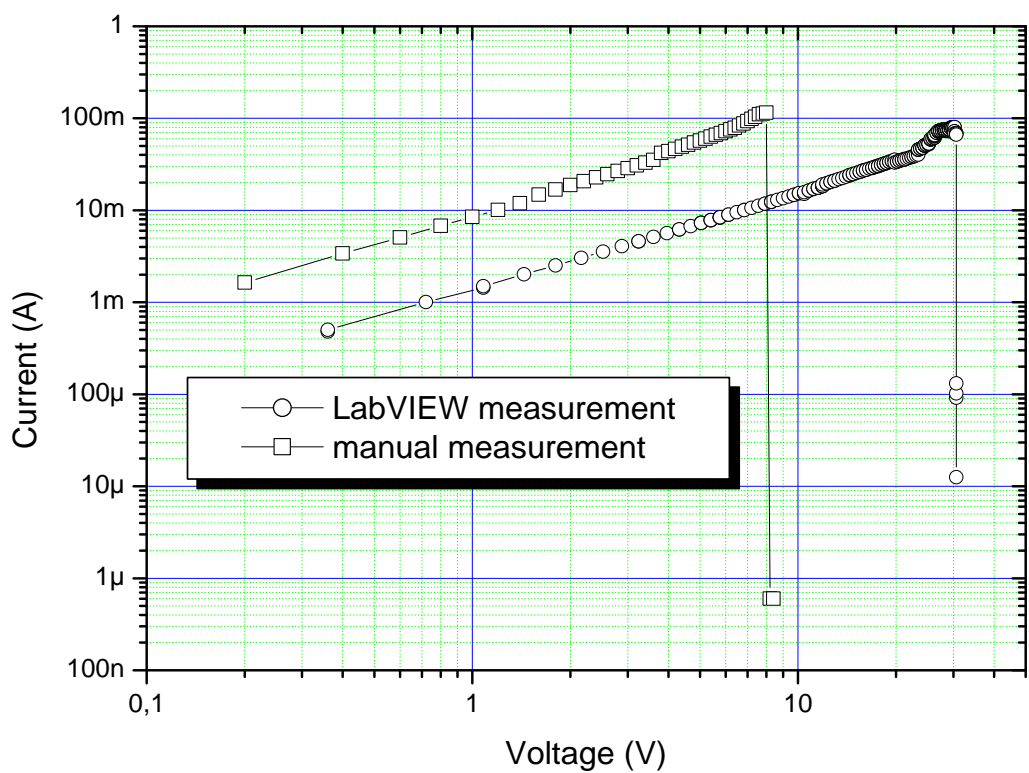

Fig.1.

More than *30* years ago such kind of behavior was observed in the glassy carbon [9] and carbon deposits [10] by famous polish scientist Kazimierz Antonowicz (1914-2002). It is turning out that electroresistance of these systems at some threshold voltage (*2-10 V*) spasmodically drops approximately on an order of magnitude. As the preparation technique of carbon films in Ref. [23] are rather close to those of [9-10], it is possible to assume a similarity of switching effects in these two cases. However there are also essential differences. At first, the jump of resistance in our experiments makes *4-5* orders of its magnitude (see Fig.1). Secondly, the resistance jump in [23] in difference with [9-10] regularly depends on temperature. Later on K.Antonowicz produce the *Al-C-Al* sandvich with thin *10*-Angstrom layer of amorphous carbon in which he observed the resistance jump up to *3* orders of magnitude and the effect of microwave detection. The later was revealed as the influence of microvawe irradiation on the transport current and were explained in terms of the Josephson effect in single Josephson contact at room temperature [12]. The resistance of the sample in Antonowicz's experiments was recovered after relaxation time of about a few days. In our experiments we also observed the relaxation times of the same order of magnitude. The reason for such relaxation is not clear so far however we have some idea about its origin, which is presented below. During long time we have not been aware of Antonowicz's results and reproduced many of his findings independently but on the base of principly another statements and fashion as it will be seen below.

    Current limiters carry out a role of protection against short circuits, which is a problem in the power systems of any class. The Josephson *FES* differs from the available analogues of household switchers and current limiters in that it is a completely electronic device in which there are no mechanical contacts, the relay, bimetallic plates and the other similar accessories, which are necessary for traditional household devices. So far the room temperature Josephson *FES* is not realised anywhere. In the *USA* the Josephson *FES* will be realised on the base of fullerenes with working temperature of *11K* [43]. Switching from a condition with high conductivity to a condition with low conductivity occurs exclusively under change of an electric field in the device, which is due to growth of a current. The current-voltage characteristic of an active element of Josephson *FES* on an initial stage of growth of a current (voltage) changes smoothly enough, as a first approximation - linearly, however at a point with critical value of a current or a voltage there is sharp strictly vertical falling of conductivity on *4-5* orders of magnitude (see Fig.1). Thus the Josephson *FES* turns to be the resistor with the resistance in tens of *MOhm*, i.e. an insulator. As it is possible to see from Fig.1 the falling of a current in a critical point occurs strictly vertically, that can be the consequence of high enough speed of switching. The direct measurement of this value however is of interest.

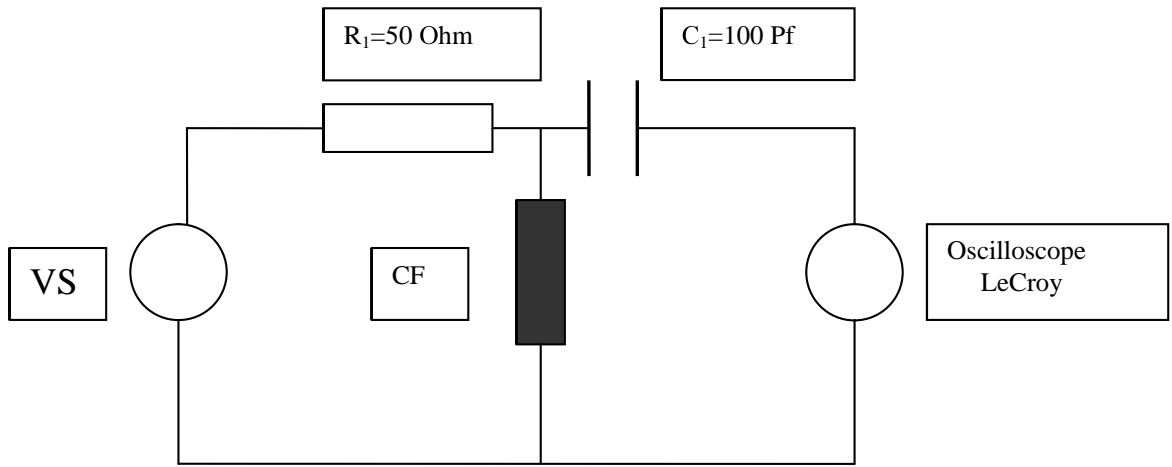

Fig.2
For the measurement of *FES* switching time the electric circuit presented in Fig.2 has been made. Here are *VS* - a voltage source, *CF* - a carbon film. Measurements were made with carbon films of two types: *CA* and chemical vapor deposition (*CVD*). A feature of the last type of film is that they need not be annealed before using, as it is the case for *CA* - films. The electric contacts to the film surface were made by both methods: mechanical and vacuum deposition. The results of measurements did not depend on the type of contacts.

As the measuring device the four-channel digital oscilloscope *LeCroy WaveRunner 6100* was used. The maximal frequency of digitization of *10 GHz* that allows measuring the processes with front width up to *225* picoseconds, and a memory size on one channel - *2 MB*. As a source of a voltage the standard block of stabilized voltage supply *TEC-9* with a limiting voltage up to *100 V* and a current up to *250 mA* was used. The electric scheme of Fig.2 was constructed so that, on the one hand, to provide some voltage drop on the active element *CF*. For this purpose the resistor $R_1$ is entered in the scheme. Thus, at catastrophic drop of current on the active element *CF*, the voltage drop on it also will sharply decrease, that will be compensated by increase in a voltage drop at the resistor $R_1$. On the other hand it is necessary to decouple on a direct current of a circuit of the voltage supply and an oscilloscope. For this purpose the capacitor $C_1$ is entered into the scheme which capacity is chosen such that the time constant of circuit $C_1$ - oscilloscope was much greater than expected times of transients in circuit *VS* - $R_1$ - *CF*.

At the measurement of parameters of transient in *CF* the circumstance is used, that at the sharp termination of a current in *CF* the potential of the left plate of capacitor $C_1$ also changes (with characteristic time of transient in circuit *CF*). That causes the recharge of capacitor and, consequently, an alternating current in a circuit of an oscilloscope. Thus, the oscilloscope, which is synchronized on single events, can define the parameters of transient, i.e. the switching time of *CF*. The results of such measurements for the film produced by *CVD* technology are presented in Fig.3. As it is possible to see from Fig.3 the time of increase of transient front makes ~*1* nanosecond. As a whole the plot in Fig. 3 reminds falling off of an alternating current in a circuit with attenuation. As it is well known that the period of oscillations of a damping alternating current only slightly differs from those without attenuation. As shown in the ref. [44] the active Josephson element can be considered as set of internal capacity and inductance so, that instead of Fig.2 we obtain the equivalent scheme of a delay line presented in Fig.4.

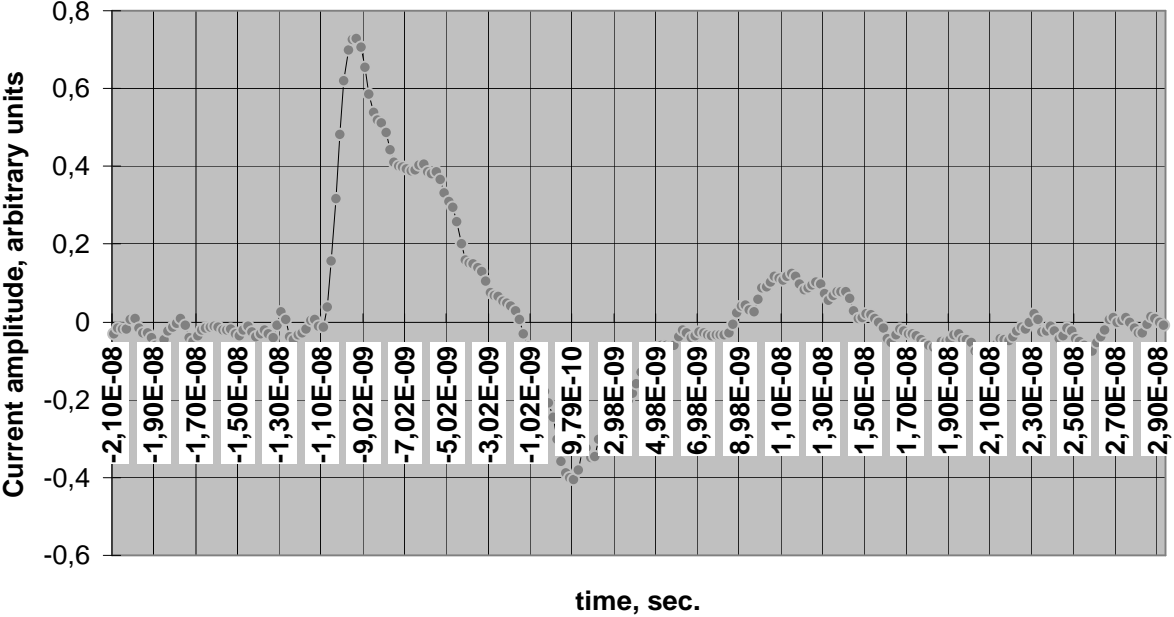

Fig.3

Internal inductance *L* and capacity $C_2$ of Josephson Element (*JE*) are rather important parameters describing its structure, geometry and electrodynamics. The opportunity of definition of values of *L* and $C_2$ through the parameters of the transient represented in Fig.3 therefore is of interest. Further we shall describe the procedure, allowing defining the specified key parameters. To define the *L* and $C_2$ we need two equations connecting these values with the data obtained during the measurement. First of such equation connects *L* and $C_2$

with the period of damping oscillations which can be directly measured from the plot of Fig.3 as follows:

$$\frac{1}{\sqrt{LC_2}} = \frac{2\pi}{T_o}, \quad (1)$$

where $T_o = 1.2 \cdot 10^{-8}$ sec is the period of oscillations of an alternating current on the plot of Fig.3. The second equation can be obtained under suggestion that $R_1C_2$ represents an integrator, with the attenuation time of transients of $3\tau = 2.5 \cdot 10^{-8}$ sec., where $\tau = R_1C_2$ is a time constant of an integrator.

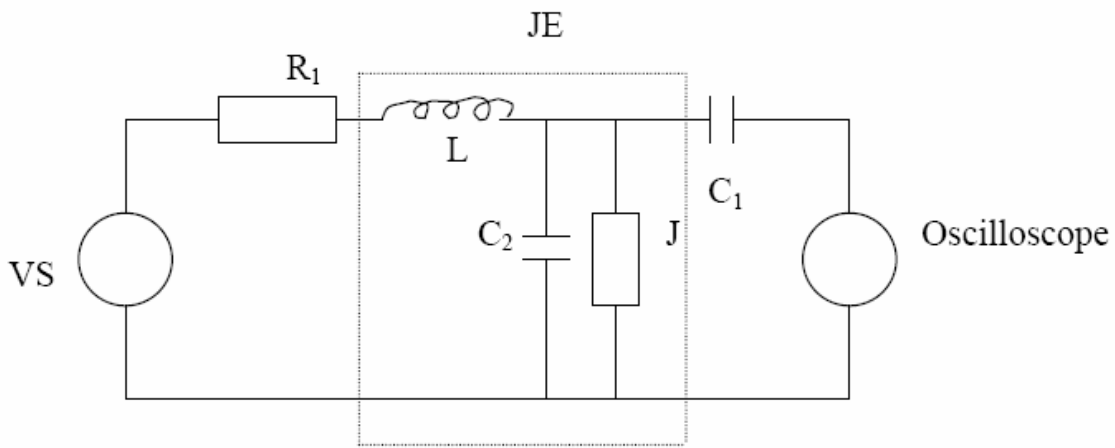

Fig.4
Whence we have obtained: $C_2 = 170$ picofarads. Then from the relation (1) we have obtained: $L_1 = 1.5 \cdot 10^{-8}$ Henri. In Fig.5 the plot of transient in the film produced by *CA* technology is presented. As it is possible to see the time of increase of front of transient in this case not strongly differs from *CVD* film and makes nearby *2* nanoseconds.

The switching time between high to low conductivity at a level of *1-2* nanoseconds does not allow speaking about the thermal mechanism of switching and, more likely, testifies about its electronic character. Certainly, on the base of data obtained so far it is impossible to make the unequivocal conclusion about the superconducting mechanism of switching, however it is necessary to note, that superconducting switches have the similar times of switching [45]. It is necessary to note, that the plots of transients shown in Figs.3 and 5 are not something exclusive; on the contrary, they are chosen arbitrary from the large number of similar results.

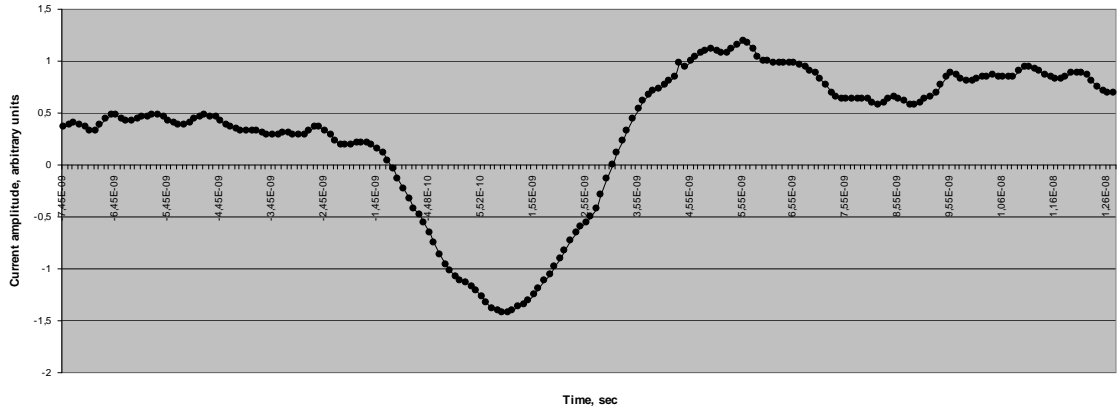

Fig.5
## B. MICROWAVE DETECTION

In order to check the existence of *JJA* in *CA* films, we performed the measurements of *RJE*. We have studied the *RJE* in various experimental arrangements of Refs. [29] and [46] and measured the induced *dc* voltage, $V_{dc}$, as a function of frequency, temperature, and input *ac* amplitude. The fluctuations of $V_{dc}$ are fast and it may be the result of a series combination of a large number of individual Josephson junctions in accordance with the CA films structure as it is described in Sect. II A.

When the structure of carbon films has been studied, it became clear, that they represent conglomerates of graphite-like granules - nanoclusters incorporated in the "matrix" of amorphous carbon [27]. Hence, the neighbouring granules divided by the isolating layer of amorphous carbon, form a *JC*. The electric properties of similar granular film very much remind the behaviour of *JM*. Therefore we have initially assumed, that such film represent a *JM*. We saw the problem in proving its presence. Now the methodology of *JM* identification is developed well enough. Successes in this direction have been appreciably reached owing to studying of new high-temperature superconductors (see the Sect. II A.), which, as it is known, represent the *JM*. It was found out, that at microwave irradiation of *JM* the constant voltage is induced, i.e. there is reversed *ac* Josephson's effect. This process reminds rectification of an alternating current, but is essentially distinct from the last as it is shown in Sect. II A.

To identify unequivocally the *JM*, it is necessary not only a technique of distinction of the reversed Josephson's effect and diodic rectification, but also rejection thermal emf and other side effects. Such technique has been developed by J.T.Chen with colleagues [29] and was described in the Sec.II.A.

J.T.Chen et al. investigated an unstable mixture of a superconducting phase with critical temperature of *240°K*, containing in the sample of *HTSC* - ceramics with the critical temperature of *90°K*. The reaction of *JM* on the microwave radiation, the dependence on temperature, frequency and amplitude of the microwave signal has been as a result thoroughly studied. Owing to this technique it was possible to prove the presence of the *HTSC* - phase with $T_c$ = *240°K*, that considerably exceeds the temperature limit reached for today of *130°K*. We have applied this technique to search of a possible *HTSC* - phase in the carbon films. During experiments all of characteristic reactions of *JM* have been reproduced and therefore its existence in a carbon film is proved. The critical temperature of the *HTSC* - phase is defined as a point where the constant microwave induced voltage $V_{dc}$ tends to be zero. The temperature dependence of $V_{dc}$ was measured by using the electrical circuit of Ref. [46] at fixed input amplitude and *RF* frequency. At all temperatures, the data were obtained after reaching thermal equilibrium. The result is shown in Fig. 6 for *300 K < T < 777 K* at $V_{ac}$ = *10 V* and *RF* frequency *f = 1 MHz*. $V_{dc}$ exponentially decreases with temperature. On the plot of dependence of $V_{dc}$ vs temperature (Fig.6) it is possible to see, that $T_c$ makes *650°K*. The observable behavior reminds very much the "hot" superconductivity with the critical temperature well above the room temperature.

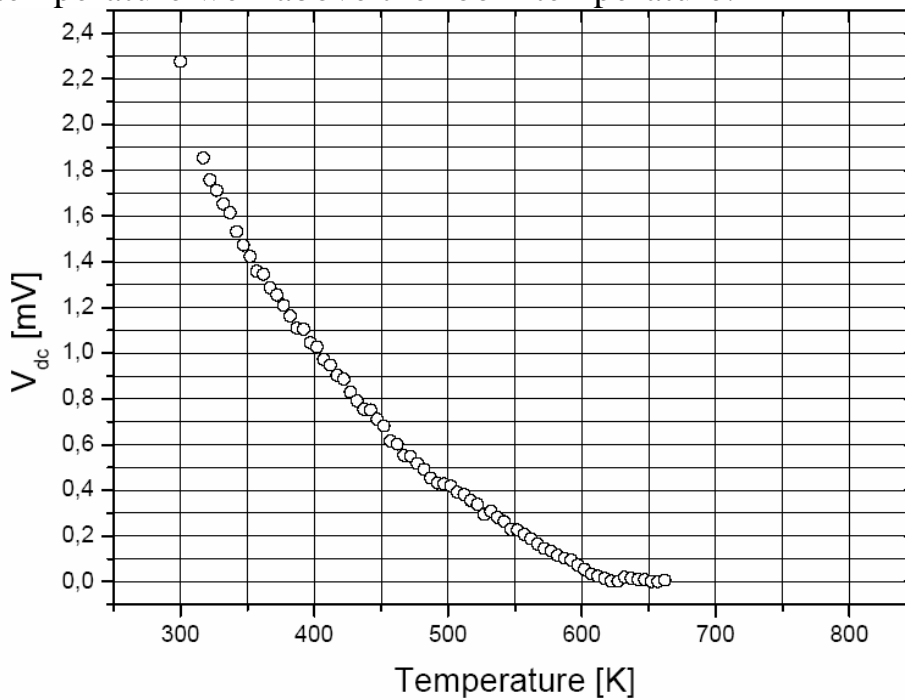

Fig.6

## C. MAGNETIC PROPERTIES

*AFM* and *MFM* measurements were employed to study the topography of film surfaces and possible magnetic clusters. Magnetic force gradient images and sample topography were obtained simultaneously with a Nanoscope III scanning probe microscope from Digital Instruments. The microscope was operated in the "tapping/lift ™" scanning mode, to separate short-range topographic effects from any long-range magnetic signal. The scanning probes were the batch of fabricated *Si* cantilevers with the pyramidal tips coated with a magnetic *CoCr* film alloy. Prior to acquiring an image, the probe was exposed to a *1-3 kOe* with the permanent magnet, which aligned its magnetization normal to the sample surface direction (i.e., *z* direction). All *MFM* data shown in this Section were collected with the tip magnetized nearly perpendicular to the sample surface, making the *MFM* sensitive to the second derivative of the *z* component of the sample stray field. Images were taken with various tip-sample separations *(10 ~ 100 nm)* with the requirement that the general shape of *MFM* images would not change with the tip-sample distance variation. This excluded any influence of the *MFM* tip on the sample micromagnetic structure and verified that any non-*z* components of the tip magnetization will contribute negligibly to the *MFM* measurements.

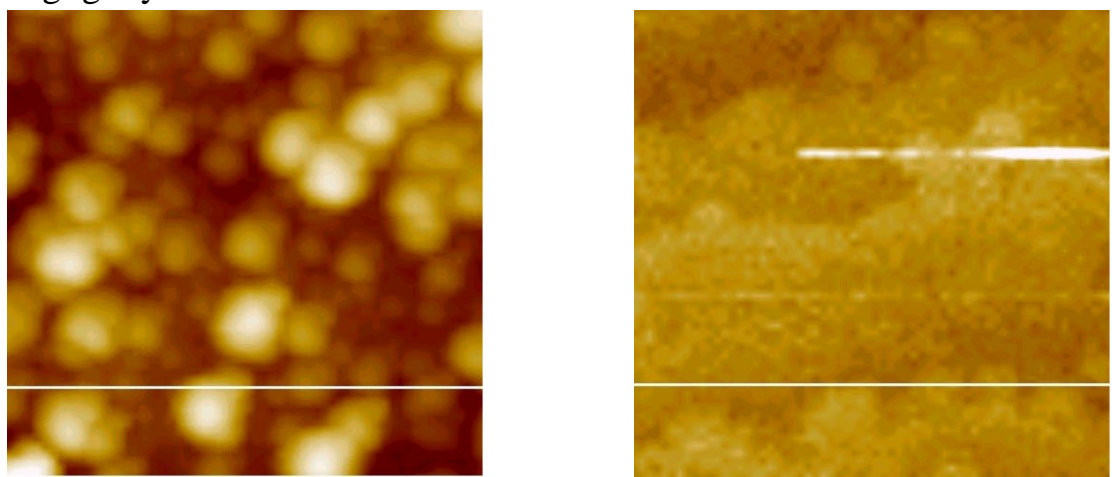

Fig.7

In fresh films there was evidence of corrugating or warping of the surface in the topographic *AFM* images and an absence of magnetic clusters or domains in *MFM* images. Here the scan size was *5 μm x 5 μm* and the scan height in *MFM* measurements was *50 nm*. The annealed films had a smooth surface at large scale and revealed the topographic clusters with the average size of about *165 nm* (see the right side of Fig. 7). As can be seen from the *MFM* image in the left side of Figure 7, it is possible to resolve the magnetic domains or particles. After magnetization of *CA* film with the use of permanent magnet with the magnetic field of *80 kamp/m* there was clear evidence of magnetic clusters with the same sizes and locations (see the right side of Fig. 8) as the topographic clusters.

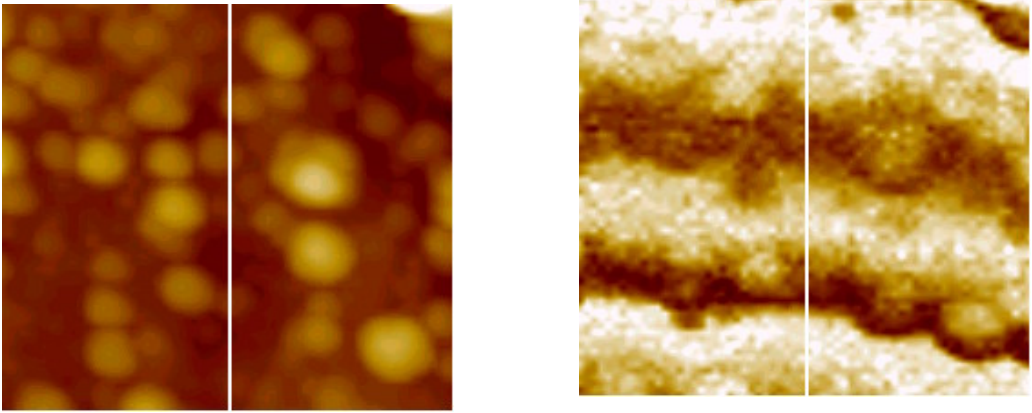

Fig.8

This suggests that there were magnetic clusters or domains in the annealed *CA* films and the magnetization of magnetic domain is small, which however can be resolved in *dc SQUID* magnetization measurements. *dc* magnetization *M(H,T)* measurements were performed with a *SQUID* magnetometer *MPMS7* from Quantum Design. The *M(H)* curves for different temperatures are shown in Fig.9. The magnetization oscillations in the magnetic fields of $10^4$-$5\cdot 10^4$ *Hauss* [26] can be seen. For clearer the picture of the oscillating behavior of magnetization under some magnification is shown in Inset of Fig.9. This picture has been obtained after subtraction of the linear diamagnetic background *M=* -$\chi(T)H$. The value of a magnetic field corresponding to jumps of magnetization, and also their amplitude depend on the temperature (see Fig.9). Each oscillation is believed to relate with the increase of a magnetic flux by one quantum $\Phi_0$ in the magnetic cluster of carbon film. Using the data of measurements, we have defined the average size of magnetic clusters being about *0.1 microns*. In the *MFM* we have seen also the magnetic clusters and we have defined their average size being about *0.16 microns*. This is good enough coincidence with the *SQUID* data.

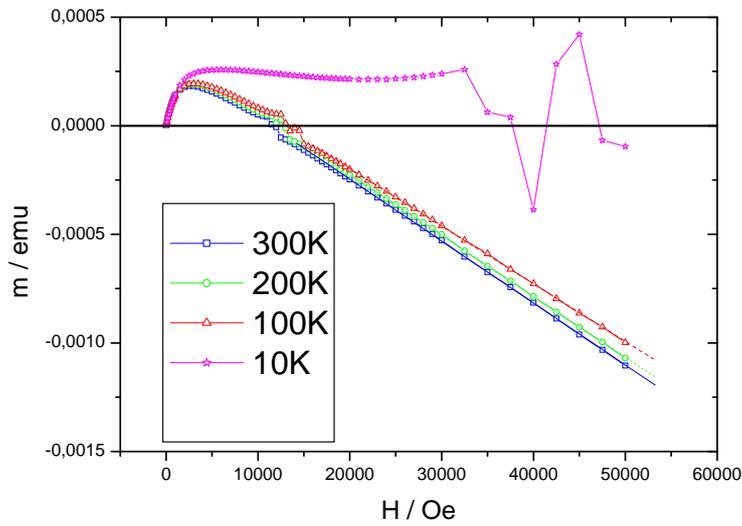

Fig.9
## D. ELECTRICAL RESISTIVITY

The temperature dependence of resistance was measured in a commercial closed cycle refrigerator below (*20 K< T <325 K*) and in a vacuum furnace and above (*300 K < T < 777 K*) room temperature, respectively. In the low temperature resistance measurements, a Lakeshore *330*-temperature controller

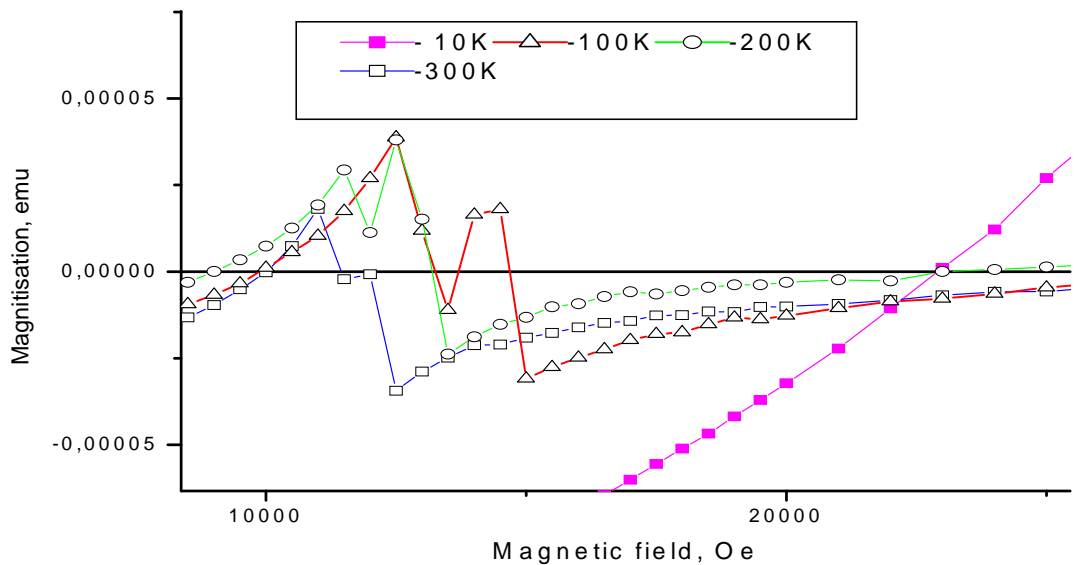

Inset of Fig.9

digitally recorded the sample temperature using a *RS*-Components *PT-100* platinum resistor placed directly on the sample. With this cryogenic system, samples could be cooled down to *20 K* in about *2 h*. Warming up of the system took longer, up to *10 h* to go from the lowest temperature to room temperature. A Keithley *220* programmable current source was used to supply a current, *I*, which was also monitored by a four-wire precision resistor of *0.1Ω*. The voltage drop, *V*, on this resistor and on the sample was measured with a Keithley *2182* Nanovoltmeter. An Agilent *34970A* Switch Unit, with a *34901A* 20 Channel Multiplexer card and a *34970A* 4×8 Matrix Switch card were used to allow for switching of both voltages and the current polarity, respectively, interconnecting the test leads with the nanovoltmeter and the current source in a break-before-make mode. All these instruments had *IEEE-488.2* interfaces and used an *IBM*-compatible personal computer as the controller. The control software, written in *LabVIEW* from National Instruments, programmed a current value and selected the current polarity between the source and the sample. Therefore both voltage drop, *V*, the sample temperature, *T*, and the time, *t*, were digitally recorded. All measurements were continuously stored in *ASCII*-files while the sample was either cooled down or warmed up.

Figure 10 shows the temperature dependence of electrical resistance *R(T)* in annealed CA films. The resistance smoothly decreases as temperature increases from *20 K* to *777 K*. This tendency is reproducible with warming up and cooling down. The change of resistance with temperature from room temperature to 777 K is about 17 ~ 25 %.

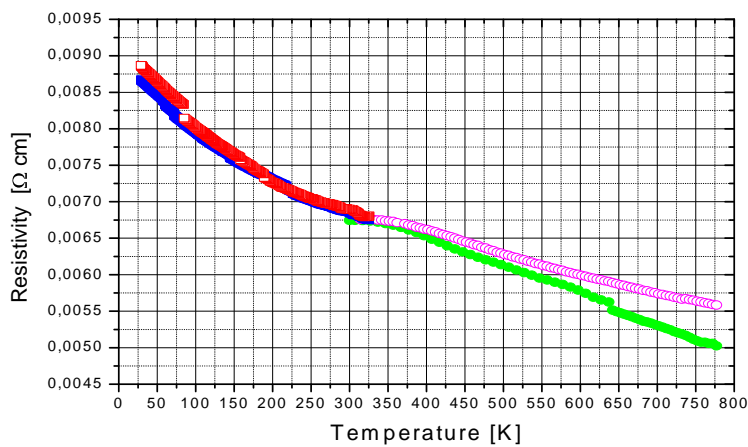

Fig.10

**E. SWITCHING INDUCED OPTICAL RADIATION**

### 1. Registration of IR radiation with the slow photodiode

As can be seen from Fig.1, during the moment of switching the power *P* developed by a current makes some Watts. At the area of a surface of the film sample of *~1 cm²* it can produces the specific power characteristic for household iron at the weigh difference in *$10^4$* times. During the moment of switching all power allocated on a linear path of current-voltage characteristic, should turn into heat or radiation. We shall estimate the change *ΔT* of temperature of a film per second till the moment of switching by the well-known relation:

$$P = mC\Delta T . \qquad (2)$$

Substituting the values of *m~200 μg* which is the weight of the film sample, *C~1000 J / kg °K* – a thermal capacity of graphite, we have obtained:

$$\Delta T = 5000\text{-}10000 \text{ °K/sec.} \qquad (3)$$

Such rate of a warming up could fuse both a carbon film, and their quartz substrate. Simple estimations show, that temperature cannot be sufficiently reduced due to heat removal by heat conductivity or air convection. In reality the temperature of a film in our experiments does not exceed *$100^oC$*. This can be explained only due to heat removal by self-radiation. All heated objects radiate photons with a spectrum of a black body. Distinction is that in the case of Josephson media radiation should be coherent and consist of one basic harmonic and several subharmonics. The main radiation frequency of Josephson media is defined by the critical temperature *$T_c$* of superconducting transition and at *$T_c$ ~ $650^oK$* makes *$10^{14}Hz$*, which corresponds to a wavelength of about *2 microns* in the *IR* band. As it has been shown in the Sect.III.A, the duration of an impulse of switching makes *1-2* nanoseconds then in the case of transformation of all power of a current into coherent radiation we have obtained the tempting prospect of creation of *IR* laser with pulse power as high as *1 GW*.

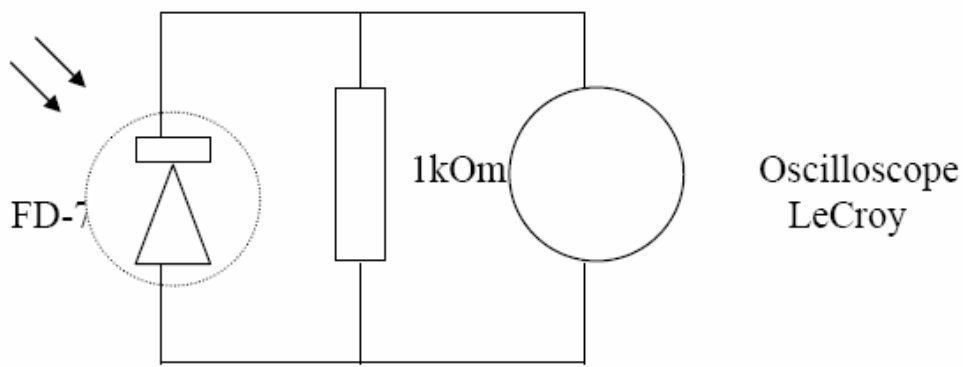

Fig.11

The simple and cheap way of registration of *IR* is the using of the photo diode. Modern photo diodes possess good sensitivity, a suitable spectral range, and *PIN* – photo diodes have high speed – some of their modifications can process the signals with the front width in hundreds of picoseconds. However as a first step we have tried to register *IR* on a linear path of current-voltage characteristic. For this purpose the domestic photo diode *FD-7* with the wide aperture of an entrance window and a maximum sensitivity on the wavelength of *0.9* microns and a high level of internal amplification of the signal was used. This allows applying the signal directly on an oscilloscope. The limiting registered wavelength of *FD-7* is bounded by the value of *1.5* microns caused by the absorption of *IR* in a glass entrance window. Speed of the diode response does not exceed *1 millisecond*; therefore with its help it is impossible to register the pulse generation. The electric scheme presented on Fig.11 has been established for the measurements.

The results of the measurement are presented in Fig.12. In Fig.12 it is possible to see a thermal impulse with duration of *3* milliseconds on which background the high-frequency generation is visible. It is known [45], when biasing with *dc* voltage *V* the Josephson's contact starts to generate the radiation with the frequency *v/V=483.6 MHz/μV*. Using this relation and the data of Fig.1, it is possible to estimate the number of Josephson's contacts as high as *~300*. From Fig.12 it is visible, that in the field of a maximum of thermal peak the amplitudes of generating pulses considerably increase, that, possibly, is connected with increasing the instability of an electrodynamics condition. Apparently, high-amplitude peaks in the area of a maximum of thermal peak reflect the energy release by means of generation of short microwave pulses, which cannot be fully resolved by photo diode *FD-7* because of its low response speed.

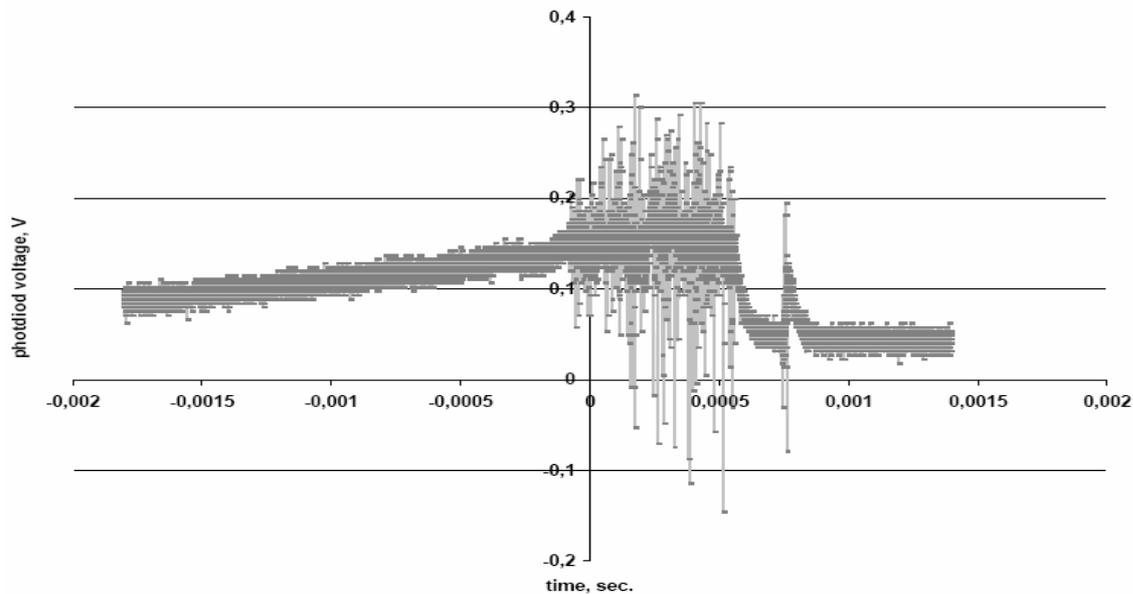

Fig.12

The given conclusion proves to be true that in a point of *0.5 ms* on the plot of Fig.12 sharp decrease of a thermal pulse is visible that, at least, can be partially caused by the microwave radiation. The basic peak of the microwave radiation should be characterized by width of some nanoseconds, which is caused by time of switching from highly conductive state (see Fig.3 and Fig.5) and cannot be registered by means of the slow photo diode.

## 2. Registration of IR radiation with the fast photodiode

For registration of *IR* radiations during the moment of switching the fast silicon avalanche photo diode have been used which is applied in fibre-optical lines of communication. These devices possess the response speed in hundreds of picoseconds and a spectrum of registration from *600* nanometers up to *1500* nanometers. The working window of the photo diode in diameter of *1.1 mm* consists of *556* pixels that enable to register radiation in a wide angular cone. Sensitivity of the photo diode makes *0.8 mV/photoelectron*, and the photoelectron is born in *90* % of cases of interaction of photons with sensitive substance of the photo diode. The scheme of connection of the photo diode is presented in Fig. 13.

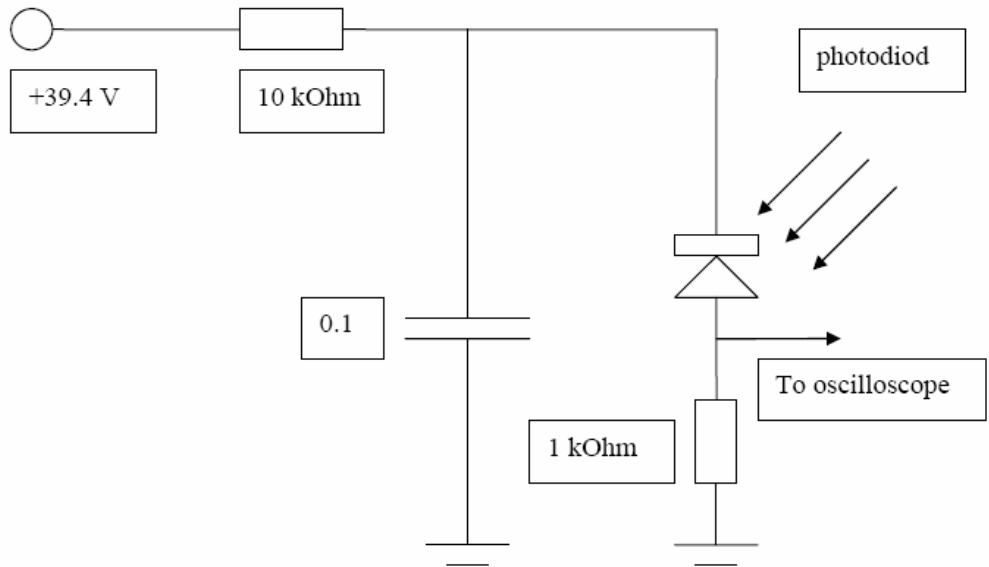

Fig.13

As the photo diode is sensitive to visible light the registration of radiation from a carbon film near to the moment of switching of conductivity was made in the blacked out chamber. The sensitive area of the photo diode settled down near to a radiating surface of a carbon film. The oscilloscope was forced to work in a mode of synchronization from single pulses. The photo diode was biased with *39.4 V* of DC voltage and its reaction to visible light was checked. At the closed cover of the blacked out chamber the background indications of an oscilloscope did not exceed *1 mV*. Then the DC voltage on contacts of a carbon film slowly increased up to critical value at which the current in a circuit of a film jumped to zero value. During the moment close to switching an oscilloscope registered the optical radiation, the plot of which time dependence is presented in Fig.14. On a background of "substrate" it is visible a series of optical pulses which amplitudes considerably exceeds a level of "substrate" so enters the photo diode into a condition of saturation. Division of process of generation into some stages can cause the presence of several consecutive optical impulses. Such representation contradicts with the data on much more transient and coherent switching of conductivity during *1* nanosecond. However as it was mentioned before the oscilloscope was synchronized on single switching events. So it is possible to suppose that each optical impulse corresponds to switching one. That is in reality in the case of optical radiation of Fig.14 the true switching picture consists of corresponding number of swithing pulses, from which the oscilloscope visualized only the first. The fact that the duration of optical pulses considerably exceeds the switching time may be explained due to limited propagation rate of optical radiation in carbon film body. Time of propagation of radiation in a carbon film with the thickness $h\sim 1\mu m$ till the moment of switching will make $\tau\sim h^2C/\lambda \sim 5\cdot 10^{-6}$ sec. The delay can increase due to decrease of heat conductivity $\lambda$, which is connected with conductivity $\sigma$ by the well-known Wideman –France law:

$$\lambda = L_0 \sigma T, \qquad (4)$$

where $L_0 = 2.445\cdot 10^8$ $Wohm/^0K^2$- is the Lorentz's number, $\sigma$- the electrical conductivity of a film, $T$ – the temperature. As can be seen from Fig.1, the conductivity of a film before switching makes $2\cdot 10^4$ $(Ohmm)^{-1}$. If we assume, that after switching the conductivity will fall in $10^4$ times as on Fig.1 then $\tau$ will make even *50 ms*. Of course there is no reason to consider the all parts of film as insulator. So the time delay can be successfully explained.

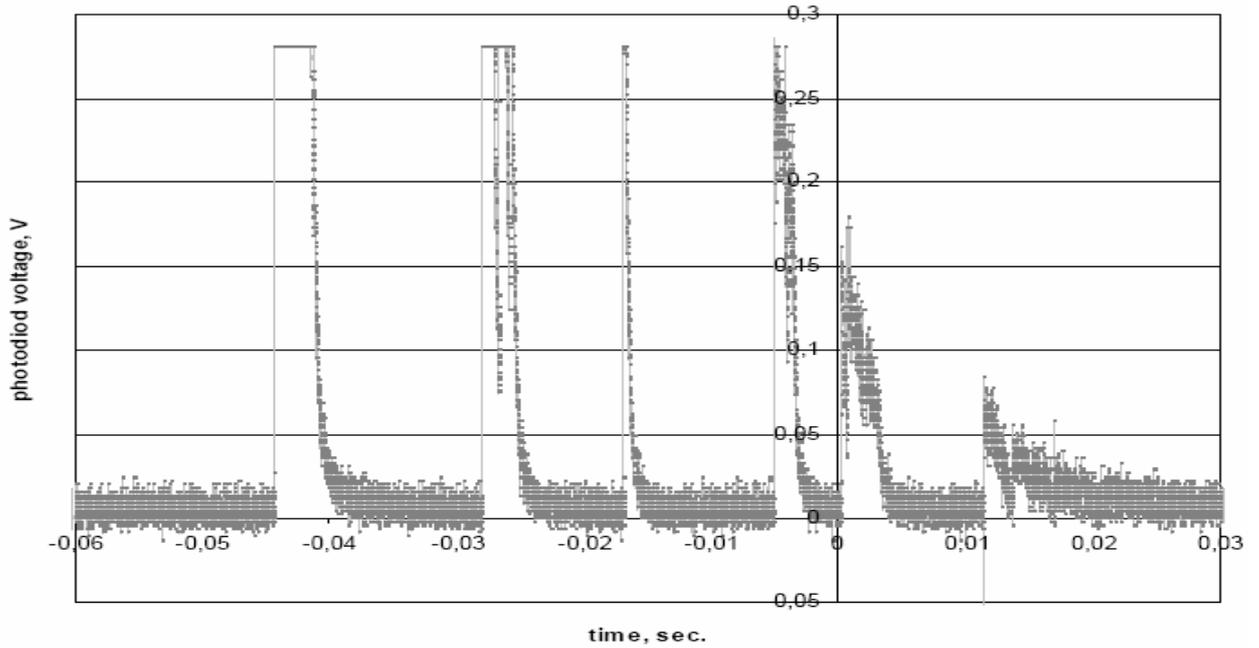

Fig.14

In fact, as it was mentioned before, in the absence of global phase coherence the conductivity of granular system has the percollation character. This means the system is composed by a number of branches of finite size conduction clusters. So the swithing process may consist from a few steps of consequent "turning off" of individual branches of conductive clusters. The first turn off happens in the branch where the current first exceeds the critical value. Then the remaining branches turn off consequently resulting at last in zero current. It is believed this picture is broadly in line with the data of Fig.14. If the motion of magnetic vortices explains the change of conductivity in carbon films, then Fig.14 displays the transformation of vortices into the *IR* radiation. This gives us some idea of a new type of laser based on the motion of magnetic vortices. In this case we can consider the Josephson media of granular structure as laser active media. Due to pinning of vortices in the film body their movement is thermally activated. That is the long relaxation time is needed for the recovery of conductivity after switching. Hovewer it is expected that under intense *IR* irradiation the relaxation time will be decreased like to pumping process in laser.

As to the wavelength of optical radiation from a carbon film it is possible to tell, that, first, it lies outside of an interval of visible light. Secondly, this length of a wave obviously is less than *1.5* microns – the limit of transmission for the silicon photo diode. Thus, it is possible to define with a high degree of probability, that the wavelength of radiation lies near to *1* micron, which agrees well with the mechanism of radiation connected with the destruction of magnetic vortices in a superconductor with the critical temperature *650-700K*.

### F.  SUMMARY OF EXPERIMENTAL RESULTS AND DISCUSSION

In Figure 10 the temperature dependence of the electrical resistance *R(T)* in the annealed *CA* films is presented. As can be seen, the resistance smoothly decreases at the increase of temperature in the range of *20 K* to *777 K*. Such resistance vs temperature behavior seems not to be a strange because many finite size loops do not produce the global phase coherence. The normal state sheet resistance of these films is about $10^7$-$10^8$ *Ohm* as can be clearly seen from Figure 1, where the current-voltage (*IV*) charactiristics of *CA* film are presented. As can be seen one of the *IV* – curve has been obtained manually and the other – by means of automatic procedure described in Section II with the Keithley *220* programmable current sources.  In Figure 1 the resistance jump up to *4-5* orders of magnitude can be seen at the critical current of about *100 mA* (at room temperature). In the frame of *SC* explanation of the effects observed the question arises: why the carbon film has a finite electrical resistance? The matter is that the superconducting systems not always can get zero resistance state or, speaking more precisely, a condition with «the global phase coherence». It becomes possible, when the resistance of a film in a normal state (e.g. at temperature above the critical) is less than the characteristic value *RQ = 7 kOhm*. However as it can be seen from Fig.1, the normal state resistance of a carbon film is the order of tens *MOhm*. Apparently, the *SC*-phase borrows only a small part of volume of the film sample that can explain its finite resistance. And whether there are the bases to assume, what superconductivity is possible in nanosized graphite granules? Apparently, the answer is yes! V.L.Ginzburg has predicted the possibility of high-temperature superconductivity in the sandwiches made of the highly conductive phase, surrounded by the material with the high dielectric permeability $\varepsilon$ [47-48]. As it was mentioned, in a carbon film the graphite granules are shipped in a matrix of amorphous carbon. By our estimations, the dielectric permeability of the carbon granular structure makes $\varepsilon = 15$ [24] (usually this value of the order of unit). And it allows considering the granular carbon film as a direct embodiment of Ginzburg idea.

B.G.Orr et al. in their works has shown [49-50] that at normal state resistance higher than RQ~*7-8 kOhm* the film have the so-called resistance reentrant behaviour. In the case of very high normal state resistance the *R(T)* curve is decreasing in all temperature range [49-50]. Some of the *CA* samples - in contrast with the common behavior of Fig.10 - showed a change of slope or even a growth of resistance in the temperature range *T >650-700 K*. However these results were generally not reproducible.

Of course more straightforward way to prove superconductivity is finding the zero-resistance state and Meissner effect in the sample. However for the graphite-like *CA* films the situation is not so lucky. As for zero-resistance, the carbon films are a composite of small graphite-like *$sp^2$*-bonding granules embedded in the matrix of amorphous carbon (see Fig.2 in Ref. [51]). Such kind of system is ruther coupled in many finite loops of granules. Many finite size loops do not produce the global phase coherence.

The dependence of $V_{dc}$ on the *RF* frequency, *f*, at room temperature is shown in Fig. 15. In this measurement, we used the same circuit as in Ref. [29]. Note that there is a polarity reversal of $V_{dc}$, which, as it is mentioned in the Sect.II.A, is the principal test, which can distinguish the *RJE* from the commonly known rectification effect. Moreover, in our *CA* films, there is a hysteresis in $V_{dc}$ with frequency as shown in Figure 15, i.e. $V_{dc}(f)$ differs depending on whether frequency was increasing or decreasing for some high frequency region i.e., *f > 9 MHz*. Such behavior also disagrees with the rectification effects.

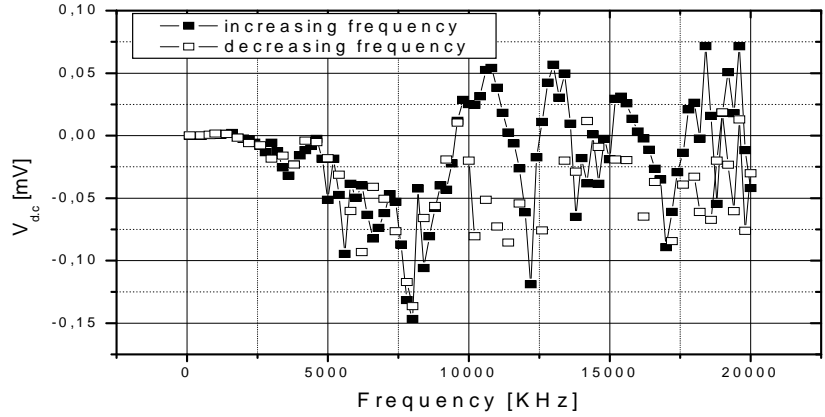

Fig.15

Figure 16 shows the change of $V_{dc}$ with the input *ac* amplitude, $V_{ac}$ for several fixed *RF* frequencies by using the same circuit as in Ref. [46]. For a small *RF* frequency region (*f < 1.6 MHz*), $V_{dc}$ increases with input *ac* amplitude and has a positive sign. However at *f = 1.6 MHz* it changes polarity (see inset of Fig. 16) and for higher frequencies (*f > 1.6 MHz*) it is negative and continues to decrease with $V_{ac}$. This is also not in line with the rectification effects.

For comparison, we studied the behavior of a standard carbon resistor, with the same resistance as the *CA* film sample. We did not observe the change of polarity in the standard carbon resistor, and the amplitude of $V_{dc}$ was one order of magnitude smaller than that for *CA* films. This reveals that the polarity change of $V_{dc}$ is a property peculiar to *CA* films.

The temperature dependence of $V_{dc}$ was measured by using the electrical circuit of Ref. [46] at fixed input amplitude and *RF* frequency. At all temperatures, the data were obtained after reaching thermal equilibrium. The result is shown in Fig. 6 for *300 K < T < 777 K* at $V_{ac}$ = *10 V* and *RF* frequency *f = 1 MHz*. $V_{dc}$ exponentially decreases with temperature. This can be successfully fitted by the form $V_{dc}=\alpha e^{bT}$ for *300 K < T < 650 K*, where *1/b= -104±60 (K)* which is in good agreement with the result of Refs. [46]. From *300 K to 777 K*, the change of $V_{dc}$ is more than *100 %* and thus larger than the *17 - 25 %* change in the electrical resistance shown in Figure 10. This excludes the possibility to explain the $V_{dc}$ change due to change of electrical resistance.

The *dc* magnetization *M(H,T)* measurements were performed with the *SQUID* magnetometer *MPMS7* from Quantum Design. The *M(H)* curves for different temperatures are shown in Fig.9. As can be seen from Fig.9 there is some paramagnetic response at *H<10000 Oe*. Due to the presence of *Fe*

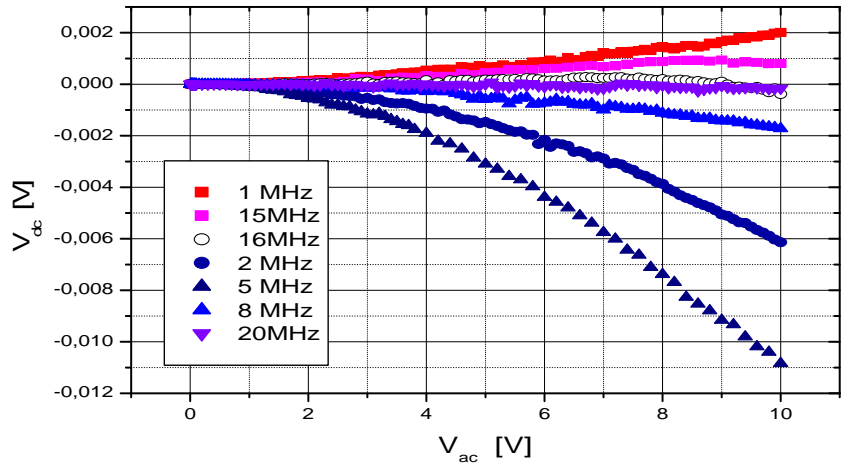

Fig.16

impurities as high as *185 ppm* the paramagnetic background signal was observed in *CA* films under *SQUID* magnetometry. Thus a strong paramagnetic background shades possible diamagnetism in *JJA* stricture. At *H>800 kamp/m* the signal is negative due to natural diamagnetism of graphite granules embedded in the matrix of both twofold and fourfold-coordinated atoms [51-52]. After subtraction of this linear diamagnetic background *M= -χ(T)H,* we found the oscillating behavior of magnetisation (see inset of Fig. 9). This is similar to the dependence of the Josephson junction critical current on a magnetic field applied in the plane of the junction (Fraunhofer pattern). As it was mentioned in the Sect.II.B, this result is in agreement with the ones obtained in Refs.[41,35,53] with only one difference - in the scale of magnetic field. The *CA* film is more comlicated system compared with one considered by C.Ebner and D.Stroud [28]. First there are Fe impurities, which produces some paramagnetic bump on the magnetization curve followed by linear decrease due to natural diamagnetizm of graphite granules. However on the background pointing above there are the ocsillations of the *dc SQUID* magnetization that is believed to be due to quantization of magnetic flux inside the loops of granules. The second difference is the temperature dependent shift of the oscillation curve not predicted by C.Ebner and D.Stroud as in their paper they "neglect the temperature dependence of the coupling". But P.Barbara and co-workers, A.P.Nielsen et al., Mahesh Chandran, C. Auletta et al., took this into account [40-42]. From Fig.9 it is visible that the first minima of magnetization occur at the value of magnetic field near *30000 Oe*. That is there is the temperature shift of ocsillation curve in agreement with the observations of Refs. [35,40-42]. As

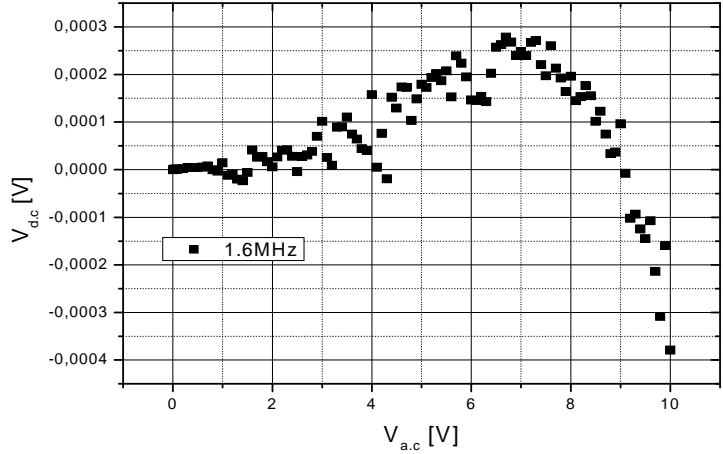

Inset of Fig.16

is well known the minima of critical current in the Fraunhofer pattern corresponds to entering of magnetic flux quanta. Therefore for the first minimum according with the results of Refs.[35,40-42] we have the relation:

$$5\Phi_0 = \mu_0 a^2 H, \qquad (1)$$

where $\Phi_0$ is the magnetic flux quantum, $\mu_0$ is the vacuum permeability, $a$ is the size of elementary *SC* loop of granules which are in the state of common phase coherence. This means that inside this elementary loop of granules the *SC* current can flow. The first maximum of *M(H)* in Fig.9 (which correspond to first minimum of critical current in a Fraunhofer pattern) gives the value of a loop size about *102 nm* as compared with the *46 μm* in the experiments described in Refs.[35,40]. According to the equation (1), this suggests a magnetic field scale $2 \times 10^5$ times larger in our case.

The data of Fig.9 also provide us with the value of critical current $I_c$ for elementary Josephson junctions composing the film granular structure. For this purpose we can use the relation from Ref.[35] which connect the magnetisation with $I_c$:

$$M = \frac{L I_c}{\mu_0 a^2} . \qquad (2)$$

Here $L$ is the inductance of the elementary loop. This gives us the value of $I_c(100K) \sim 0.6\ \mu A$ for the annealed film. That is the upper evaluation of the critical current density will be about $10^4 A/cm^2$. However taking into account the cross section of current flow between adjacent loops must be substantially smaller than the loop square we can obtain the current density close to the pairs breaking value.

The oscillations of magnetization in *CA* films differ from that observed on the model *JJA* structure of [35,41-42] in the scale of magnetic field. In our case the magnetic field where the oscillation starts is about *1T* as compared with *50 mOe* in the experiments of Refs. [35,41-42]. However the *MFM* study reveals the size of elementary conductive loop of granules in *CA* films being about *500* times smaller than in the model *JJA* structure yet having the same magnetic flux per *SC* loop. So it is believed we can proceed our measurements and analysis partly in a fashion similar to Refs. [35,41-42].

Comparing the *MFM* and *AFM* pictures, we have noticed, that; at least, some magnetic clusters coincide with the topological ones (see Fig.7 and Fig.8). The difference between Figs. 7 and 8 is that in the later case some preliminary magnetization of the film by the permanent magnet with the field of about *0.1 T* has been made. As can be seen from the comparison of Figs. 7 and 8 the magnetization gives rise to considerable strengthening of the magnetic clusters. The well-distinguished currents in Fig.8 flow around the borders of clusters, which remind the magnetic vortices. Whether the clusters visible in a *MFM* are the magnetic vortices? It is believed this question will be answered in the new cycle of magnetic measurements.

The size of the elementary loop inferred from the magnetisation experiments is nearly the same as one observed in the *MFM* –measurements. It is believed this coincidence is not a casual but is, in fact, the evidence of the existence of *SC* loops in the *CA* film samples.

Other perspective direction – is the study of optical radiation emitting by a film during the moment of switching from a condition with high conductivity in a condition with high resistance (see the Sec. III. E). In fact if we deal with the *JM*, at the breaking of superconductivity it is necessary to expect the Josephson's radiation with the characteristic frequency defined by Cooper's pairs coupling energy. For a superconductor with the critical temperature of *650°K* it is necessary to search for infrared radiation with the wavelength of some micrometers. The first attempts to register of such kind radiation by means of the fast high-sensitivity photodiode have appeared encouraging: it was possible to fix a series of optical impulses in the expected range, i.e. in the time vicinity near to spasmodic change of conductivity [54-55] (see Fig.6). The impulse amplitudes considerably exceed a level of "substrate" and the photo diode comes into a condition of saturation. However the duration of impulses essentially exceeds the expected value of *1* nanosecond. Apparently the existence of few impulses instead of one can be explained in terms of "multistage" generation caused by the gradual "deenergizing" of separate superconducting clusters. In that case the switching also should be step-by-step, and the general duration of the processes of switching and radiation can quite make the *100-th* fraction of second (see Fig.14). The spreading of an impulse of radiation relatively of initial nanosecond width of the switching impulse should be connected with time spent on the «brightening» and propagation processes of an optical impulse through a film and substrate body.

The observation of *IR-* radiation emitted from the granular carbon film under switchng of electrical resistance (see Figs. 12 and 14) can be explained also in terms of *HTSC*. These results can be used for development of new type of microlasers with active midia composed of magnetic vortices.

Of course the results obtained cannot be considered as unambiguous evidence of a *SC* phase or correlation in *CA* and *CVD* films. On the contrary this is only the first indication of some features in electromagnetic behavior of *CA* films, which require further investigations to check the results using other equipment and methods. One of such check method may be the study of the influence of energetic particle irradiation on the parameters measured experimentally. These could be the location and amplitude of Fraunhofer-like oscillation of magnetisation, *MFM* pictures, *RJE-* related and *R(T)* data. As is well known [56] the energetic particle irradiation gives rise to drastic modification of solid-state material structure. This is not only due to point defect generation but also due to displacement cascades occurring into the target media under energetic particles bombardment. It was shown that the cascade core is a vacancy-enriched zone while the cascade periphery is, on the contrary, an interstitial-enriched zone. With rising particle energy the size of cascade zone increases. Depending on particle energy, charge and mass the size of cascade zone may extend from a few up to hundreds and thousands atomic sites in the crystalline lattice. So cascades can modify the solid-state structure on both short and long ranges. This suggests that *CA* film structure can perhaps be changed controllably by varying particle irradiation parameters such as irradiation dose, charge, or mass and energy of incident particle. Of course such irradiation must be combined with measurement of the main electromagnetic parameters discussed above. Particularly it will be very interesting to observe the shift of Fraunhofer-like oscillations position and change of their amplitude due to, for example, neutron irradiation. Some description of such an experiment with conventional *HTSC* published in Ref. [57] shows that neutron irradiation up to *0.1 DPA ($10^{19}n/cm^2$)* gave rise to substantial change of critical current value. Irradiation with a proton beam at the Moscow Meson Facility [56] with today's energy about *300 MeV* and proton intensity up to *150 μA* can even substantially accelerate this experiment.

## G.  POSSIBLE APPLICATIONS

In the further researches we assign the greater hopes on doping of carbon films with the purpose to increase of their conductivity. Probably, it will allow reaching a zero resistance state at room temperature.

We will continue to investigate the *JM* in the granular carbon films and simultaneously we will concentrate on the possible applications of the results already obtained. The existence of *JM* at room temperature opens the prospect to develop the various kind of devices of noncryogenic Josephson's electronics [58-60].

One of similar application – is the contactless field effect switcher (*FES*) (see Fig.1). Integration of such switchers into the industry will essentially raise the safety in the electric circuits and their noise immunity. In the *USA* the similar field switcher on the base of fullerenes [43] for a long time is under development, however it requires the cooling by the liquid helium.

Other interesting application can be the Josephson's detector of γ-radiation for registration of neutrinos and a dark matter [61-62]. In their recent work [62-63] Christian Beck and Michael C. Mackey proposed to extract the density of dark energy in the Universe from the measured spectra density of current noise in the Josephson junction array. In agreement with today's theory the dark energy density $R_{dark}$ of the order *4 GeV/m³* gives rise to the upper cutoff frequency in the noise spectra of *JJA* of about *1.7 10¹² Hz*, which increases proportionally with the critical temperature of *JJA* structure. Figure 6 gives us the idea to widen the range of measurements of noise spectra with the *JJA* of granular carbon films. As can be seen from Fig.6 the critical temperature of the granular carbon film sample can reach the value of *650 K*, which can help to widen the accessible range of noise spectrum.

Some measures on improving the *CA* and *CVD* films structure are now in progress: the exclusion of paramagnetic contamination from *CA* films, enhancing the interaction between granules by means of intercalation procedure [24], probing the new techniques for carbon films production.

It is possible to create the generators and detectors of the microwave radiation. However first of all it is necessary to analyse, whether this innovations will have any advantages in comparison with the existing devices.

It seems to be interesting to use the granular carbon films as the magnetic protection. Influence of an electric current destroys the magnetic vortices in a film. Returning in highly conductive state occurs at "pumping" of vorticess from the outside with "absorption" of magnetic fields from surrounding space. Covering with a carbon film of the walls, ceiling and a floor of rooms with the strong requirements on the magnetic protection would allow supervising and preventing of penetration of magnetic fields by measuring the electrical resistance of separate sites of a coating.

Optical radiation of films at spasmodic switching of conductivity opens one more opportunity – the designing of new type of lasers based on the movement of magnetic vortices. The power of the similar laser can appear to be rather significant. As can be seen from our experiments, during the moment of switching the power developed by a current, makes some Watts. As a result of jump the resistance of a film during *1-nanosecond* increases in *10* thousand times, i.e. the film turns to be practically an insulator. Apparently, all accumulated power will be brighten in the form of several impulses of radiation with the duration in *1* nanosecond each, and we can obtain for such kind of laser on the magnetic vortices the pulse power of the order *1 GW*.

The next interesting application can be a covering with a carbon film of an internal surface of resonators for accelerators of elementary particles, and also the microwave transportation systems for the electric power with the purpose of reduction of losses, and creation of *HTSC* - wires for powerful superconducting noncryogenic magnets. The deposition technology of carbon film coverings easily allows making it from a gaseous phase (*CVD*-technology). Moreover, this process without any problems can be automated, supervising quality of a covering through the measuring of their electroresistance.

All listed things allow hoping, that for the granular carbon films the light future in power and electronics can be predicted.

## IV. CONCLUSION

- Experimental results presented in this work have been obtained with the help of such techniques as *dc SQUID* magnetisation, *MFM*, *RJE*, and *R(T)* measurements. These results support the existence of *SC* phase or fluctuation in *CA* films at room and possibly higher temperatures.

- From the *dc* magnetisation measurements the size of an elementary *SC* loop of *102 nm* has been deduced. This value has been found to be in agreement with *MFM* measurements. The critical current in such loop of *0.6 μA* has been obtained.
- The observation of *dc* voltage induced in samples due to *RJE* and their frequency and temperature dependencies, which agree well with the previous observations, has been considered as independent proof of existence of *JJA* and therefore the presence of *SC* phase or fluctuation inside *CA* films.
- The disappearance or change of polarity of *RJE* signal above *650 K* reveals that *SC* phase can persists up to *650 K*.
- Energetic neutron or even proton irradiation can help to check the results obtained and shed new light on the origin of electromagnetic features observed in *CA* films.
- Independent of the explanation of the observed results, *CA* films can be used in applications as a Josephson-like non-cryogenic devices

## ACKNOWLEDGMENTS

I'm gratefully acknowledging the financial support from the Russian Foundation of Basic Research through Grant No. 05-08-17909-a.

**Figure captions**

Fig.1 Current-voltage characteristics of *FES* based on *CA* film. Two variants of switching are represented with manual and authomatic current feeding on the sample. Switching time is about *1* nanosecond. The ratio of conductivities before and after switching is about *$10^4$-$10^5$*.

Fig.2. The electric scheme for measurement of switching rate *FES* based on the granular carbon film.

Fig.3. The time dependence of transient current in carbon *CVD* film. The swithing time is defined as the time to first maximum of current.

Fig.4. The equivalent electric scheme for measurement of transient parameters under switching.

Fig.5. The time dependence of transient current in *CA* film. Again like Fig.3 the swithing time is defined by a point of first extremum.

Fig.6. Temperature dependence of the *dc* voltage induced by microwave radiation. The voltage start to be zero at *650K*.

Fig.7. The simultaneous pictures of a surface of carbon film in *AFM* (left) and *MFM* (right). Some bright spots in the AFM picture are the topographic inhomogeneity of the film surface. The weak magnetic clusters can be seen in the *MFM* picture.

Fig.8. The simultaneous images of a surface of a carbon film in *AFM* (on the left) and *MFM* (on the right) after magnetization by a permanent

magnet with field of about *0.1 T*. Correlations in the positions, of at least, some magnetic and topological clusters are visible. Along borders of the magnetic clusters the circulating currents are seen. On a photo this currents look like as dark strips along a cluster's chain and remind huge boulders in the mountain river if to look onto it from the height of the bird's flight.

Fig.9. The magnetization oscillations of the sample depending on the enclosed magnetic field for different temperatures *10, 100, 200* and *300 °K*. The form of the curve is caused by action of three factors: the primary "hill" is caused by the presence of ferromagnetic impurities, then due to natural diamagnetism of graphite magnetization linearly decreases, and on this background the spasmodic oscillations arise, which are connected with the entering of a magnetic flux quanta in the magnetic clusters of granular carbon film.

Inset of Fig.9. The ocsillations area «under a microscope». It is visible, that oscillation peaks are shifted down to lower magnetic field, and their amplitude decrease with the growth of temperature.

Fig.10. Temperature dependence of electrical resistance in the annealed *CA* film.

Fig.11. The scheme of *IR* registration with the slow photodiod *FD-7* on the linear path of current-voltage characteristic of a carbon film.

Fig. 12. Results of registration of radiation from *CVD* carbon film in a stationary mode by means of photo diode *FD-7*.

Fig.13. The scheme of connection of the high-speed photo diode.

Fig.14. The optical radiation registered by the fast photo diode from a *CVD*-carbon film near to the moment of switching.

Fig.15. Frequency dependence of $V_{dc}$ of *CA* films in the circuit of Ref. [29].

Fig.16. Change of induced $V_{dc}$ with the input *ac* amplitude, $V_{ac}$ for several fixed *RF* frequencies in the circuit of Ref. [46].

Inset of Fig.16. Change of polarity of $V_{dc}$ ($V_{ac}$) for *f = 1.6 MHz*.